\documentclass[english,aps,preprint]{revtex4}
\usepackage[T1]{fontenc}
\usepackage[utf8]{inputenc}
\usepackage{textcomp}
\usepackage{amstext}
\usepackage{graphicx}
\usepackage{esint}

\makeatletter
\@ifundefined{textcolor}{}
{%
 \definecolor{BLACK}{gray}{0}
 \definecolor{WHITE}{gray}{1}
 \definecolor{RED}{rgb}{1,0,0}
 \definecolor{GREEN}{rgb}{0,1,0}
 \definecolor{BLUE}{rgb}{0,0,1}
 \definecolor{CYAN}{cmyk}{1,0,0,0}
 \definecolor{MAGENTA}{cmyk}{0,1,0,0}
 \definecolor{YELLOW}{cmyk}{0,0,1,0}
 }

\@ifundefined{definecolor}
 {\usepackage{color}}{}
\makeatother

\makeatother

\usepackage{babel}

\begin{document}

\title{Revealing the electronic structure of a carbon nanotube carrying
a supercurrent}

\author{J.-D. Pillet$^{a}$, C. H. L. Quay$^{a}$, P. Morfin$^{b}$, C. Bena$^{c,d}$,
A. Levy Yeyati$^{e}$ and P. Joyez$^{a}$}

\affiliation{$^{a}$Quantronics Group, Service de Physique de l'Etat Condensé,
CNRS URA 2426, IRAMIS, CEA\\
 F-91191 Gif-sur-Yvette, France}

\affiliation{$^{b}$Laboratoire Pierre Aigrain (LPA), CNRS UMR 8551,\\
 Université Pierre et Marie Curie (Paris VI) - Université Denis
Diderot (Paris VII) - Ecole Normale Supérieure de Paris (ENS Paris)}

\affiliation{$^{c}$Laboratoire de Physique des Solides, CNRS UMR 8502, Université
Paris-Sud, Bât. 510, F-91405 Orsay, France}

\affiliation{$^{d}$Institut de Physique Théorique, CEA/Saclay, CNRS URA 2306,
Orme des Merisiers, F-91191 Gif-sur-Yvette, France}

\affiliation{$^{e}$Departamento de Física Téorica de la Materia Condensada C-V,
Universidad Autónoma de Madrid, E-28049 Madrid, Spain.}

\maketitle
\textbf{Carbon nanotubes (CNTs) are not intrinsically superconducting
but they can carry a supercurrent when connected to superconducting
electrodes \cite{kasumov1999,jarillo-herrero2006,cleuziou2006,pallecchi2008}.
This supercurrent is mainly transmitted by discrete entangled electron-hole
states confined to the nanotube, called Andreev Bound States (ABS).
These states are a key concept in mesoscopic superconductivity as
they provide a universal description of Josephson-like effects in
quantum-coherent nanostructures (e.g. molecules, nanowires, magnetic
or normal metallic layers) connected to superconducting leads \cite{beenakker2004}.
We report here the first tunneling spectroscopy of individually resolved
ABS, in a nanotube-superconductor device. Analyzing the evolution
of the ABS spectrum with a gate voltage, we show that the ABS arise
from the discrete electronic levels of the molecule and that they
reveal detailed information about the energies of these levels, their
relative spin orientation and the coupling to the leads. Such measurements
hence constitute a powerful new spectroscopic technique capable of
elucidating the electronic structure of CNT-based devices, including
those with well-coupled leads. This is relevant for conventional applications
(e.g. superconducting or normal transistors, SQUIDs \cite{cleuziou2006})
and quantum information processing (e.g. entangled electron pairs
generation \cite{recher2001,herrmann2010}, ABS-based qubits \cite{zazunov2003}
). Finally, our device is a new type of dc-measurable SQUID.}

First envisioned four decades ago \cite{kulik1970}, ABS are electronic
analogues of the resonant states in a Fabry-Pérot resonator. The cavity
is here a nanostructure and its interfaces with superconducting leads
play the role of the mirrors. Furthermore, these {}``mirrors” behave
similarly to optical phase-conjugate mirrors: because of the superconducting
pairing, electrons in the nanostructure with energies below the superconducting
gap are reflected as their time-reversed particle – a process known
as Andreev Reflection (AR). As a result, the resonant standing waves
– the ABS – are entangled pair of time-reversed electronic states
which have opposite spins (Fig. 1a); they form a set of discrete levels
within the superconducting gap (Fig. 1b) and have fermionic character.
Changing the superconducting phase difference $\varphi$ between the
leads is analoguous to moving the mirrors and changes the energies
$E_{n}(\varphi)$ of the ABS. In response, a populated ABS carries
a supercurrent $\frac{2e}{h}\frac{\partial E_{n}(\varphi)}{\partial\varphi}$
through the device, while states in the continuous spectrum (outside
the superconducting gap) have negligible or minor contributions in
most common cases \cite{beenakker2004}. Therefore, the finite set
of ABS generically determines Josephson-like effects in such systems.
As such, ABS play a central role in mesoscopic superconductivity and
can be seen as the superconducting counterpart of the Landauer channels
for the normal state: in both cases, only a handful of them suffices
to account for all the transport properties of complex many-electron
systems such as atomic contacts or CNTs. In effect, the ABS concept
quantitatively explains the Josephson effect in atomic contacts \cite{rocca2007};
it also explains tunneling spectroscopy of vortex cores and surface
states in some superconductors \cite{fischer2007}. However, there
has been to date no detailed direct spectroscopic observation of individual
ABS. Interest in such spectroscopy has increased with recent proposals
for using ABS as quantum bits \cite{zazunov2003}, and AR as a source
of entangled spin states \cite{recher2001}.

Nanotubes are particularly good candidates for the observation of
ABS. First, CNT-superconductor hybrid systems are expected to show
a small number of ABS, and the typical meV energy scales involved
in nanotube devices are comparable with conventional superconducting
gaps. These are favourable conditions for a well-resolved spectroscopy
experiment. Second, given the length of CNTs, it is possible to introduce
an additional tunnel probe which enables straightforward tunneling
spectroscopy \cite{chen2009}. Furthermore, CNTs are of fundamental
interest as nearly ideal, tunable one-dimensional systems in which
a wealth of phenomena (e.g. Luttinger-liquid behavior \cite{bockrath1999},
Kondo effects \cite{buitelaar2002,cleuziou2006} and spin-orbit coupling
\cite{kuemmeth2008}) has been observed, whose rich interplay with
superconducting coupling has attracted a lot of interest \cite{glazman1989,spivak1991,fazio1995,caux2002,dellanna2007,dolcini2008,zazunov2009}.

\begin{figure}
\includegraphics[width=12.5cm]{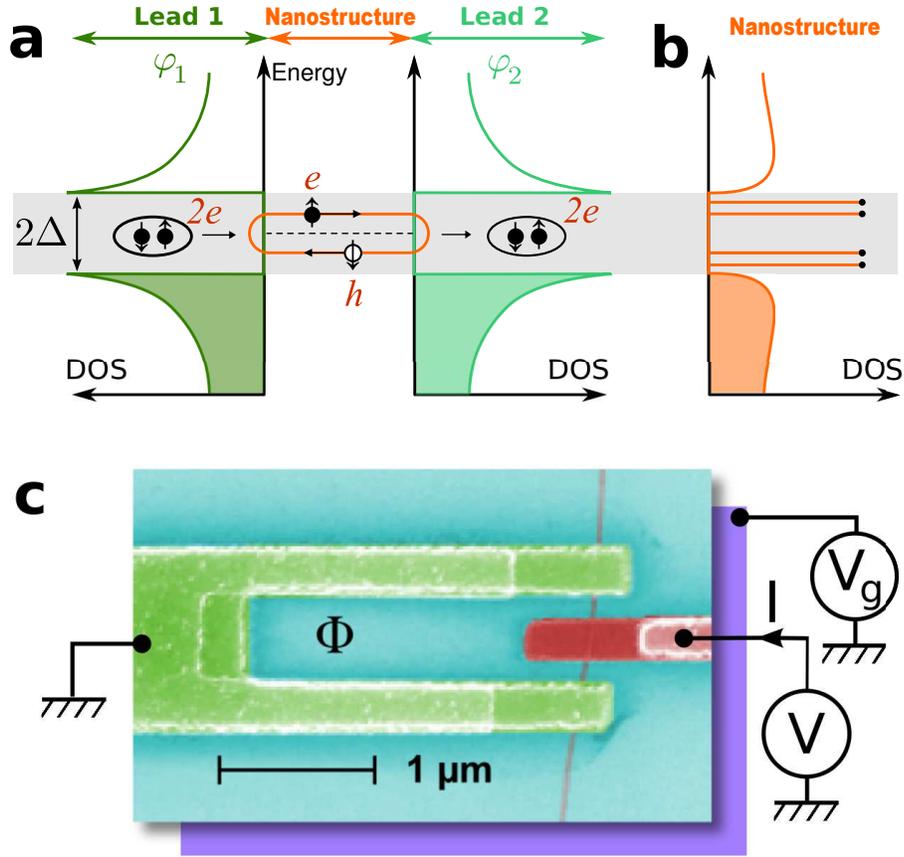}\caption{\textbf{\footnotesize a}{\footnotesize{} : Generic schematic for an
Andreev Bound State (ABS) in a nanostructure between two superconducting
leads, which have Densities of States (DOS) with a gap $\Delta$,
and with respective superconducting phases $\varphi_{1,2}$. At energies
within the superconducting gap (grey band) the Andreev reflection
process (which reflects an electron $(e)$ as a hole $(h)$ – its
time-reversed particle – and vice versa) leads to the formation of
discrete resonant states of entangled $e-h$ pairs confined between
the superconductors. These states –the ABS– are electronic analogues
to the resonances in an optical Fabry-Pérot cavity. }\textbf{\footnotesize b}{\footnotesize{}
: The local DOS in the nanostructure is thus expected to display a
set of resonances in the gap at the energies of the ABS. The energies
of the ABS should depend periodically on the superconducting phase
difference $\varphi=\varphi_{1}-\varphi_{2}$ which is analogous to
the optical cavity length.}\textbf{\footnotesize{} c}{\footnotesize{}
: Color-enhanced scanning electron micrograph of a device fabricated
for the spectroscopy of ABS in a CNT which appears here as the thin
vertical grey line. The substrate consists of highly doped silicon
serving as a back gate (figured here in violet), with a 1\textmu{}m-thick
surface oxide layer. A grounded superconducting fork (green) is well
connected to the tube, forming a loop. The measurement of the differential
conductance $\partial I/\partial V$ of a superconducting tunnel probe
(red) weakly connected to the tube gives acces to the density of states
in the CNT, where ABS are confined. The energies of the ABS can be
tuned by varying the gate voltage $V_{g}$ and the magnetic flux $\Phi$
threading the loop.}}

\end{figure}

\begin{figure}
\includegraphics[width=10cm]{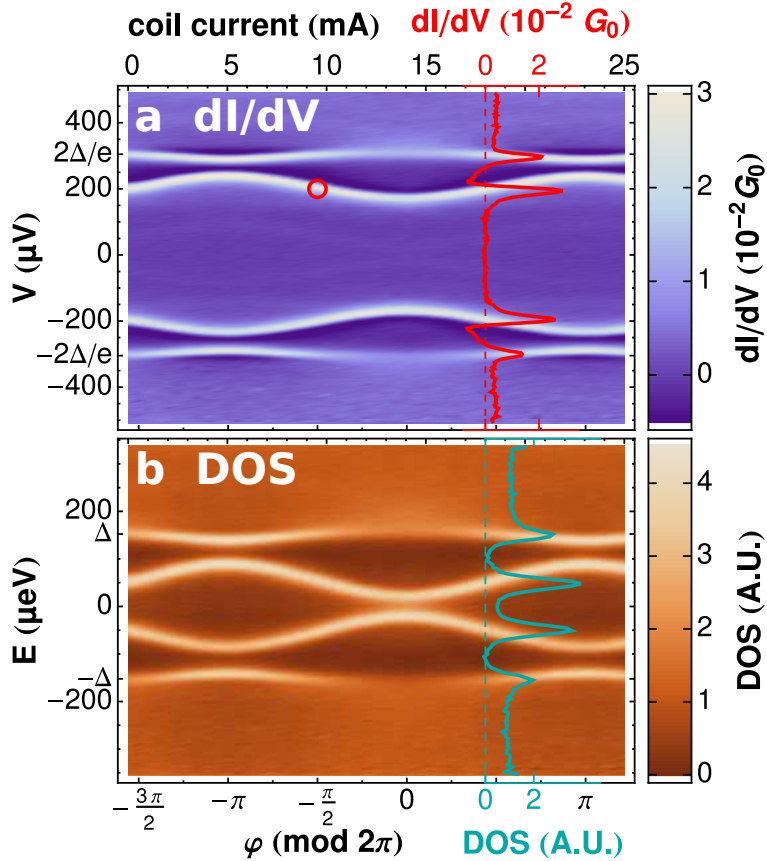}\caption{\textbf{\footnotesize a}{\footnotesize : Differential conductance
of the tunnel probe at a fixed gate voltage$V_{g}=-0.5\,\mathrm{V}$
as a function of the bias voltage $V$ of the probe junction (vertical
axis) and of the current in a coil (top axis) which controls the flux
$\Phi$ through the loop. The sharp resonances are the signature of
the ABS, and the periodicity of the pattern demonstrates that ABS
coherently connect the two end contacts and are sensitive to their
superconducting phase difference $\varphi$ (bottom axis). The solid
color traces correspond to cross sections of the data at the flux
indicated by the dashed line. $G_{0}=2e\text{\texttwosuperior}/h$
denotes the conductance quantum. }\textbf{\footnotesize b}{\footnotesize{}
DOS in the CNT as \textcompwordmark{}deconvolved from the data in
panel a, assuming BCS DOS in the tunnel probe. The device can be operated
as a dc-current SQUID magnetometer by biasing it at a point which
maximize }$\partial I/\partial\Phi${\footnotesize , as indicated
by a red circle. The fact that the phase is not zero at zero current
in the coil is due to a residual magnetic field in our setup.}}

\end{figure}

\begin{figure}
\includegraphics[width=12.5cm]{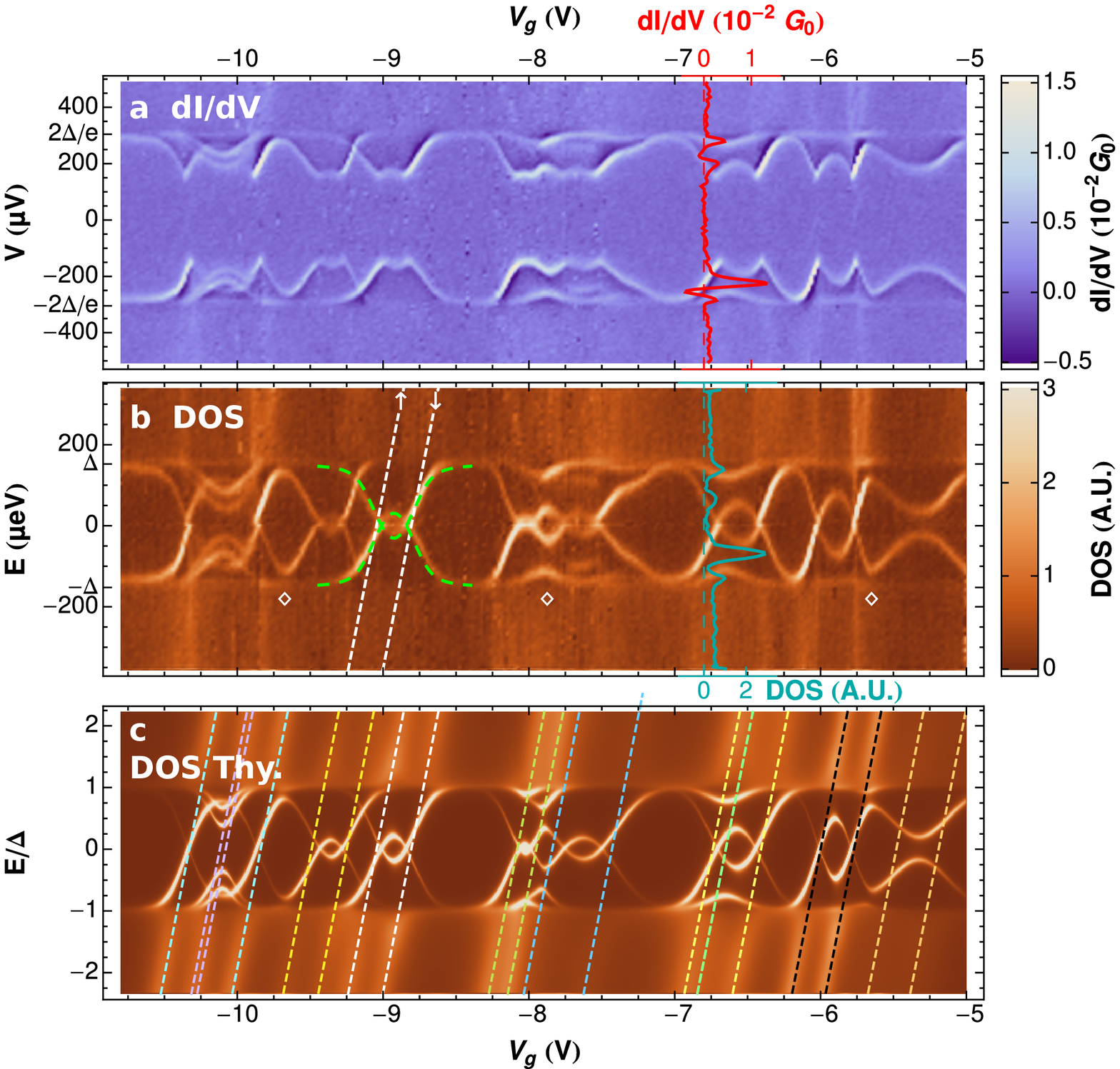}\caption{\textbf{\footnotesize Note: These color plots display well on computer
screens. However, the printed appearance might be too dark depending
on the printer used.}\protect \\
\textbf{\footnotesize a}{\footnotesize{} Gate dependence of the
differential conductance of the tunnel probe. }\textbf{\footnotesize b}{\footnotesize{}
DOS in the CNT as deconvolved from the data in panel a, after correcting
for the gating effect of the probe junction which appears as a slight
horizontal shear in panel a. The ABS form an intricate pattern of
intertwined lines. Predicted from a basic quantum-dot model for the
CNT, the green dotted bell-shaped lines are the positions of ABS arising
from a single Spin-Split Pair of Levels (SSPL - white dashed lines)
crossing the gap as the gate voltage is increased. The spin labelling
indicates only the relative orientations of the spins in these levels.
Most of the resonances observed in this panel have similar shapes
and can be attributed to different SSPL. However, some resonances
corresponding to two different SSPL are connected together where indicated
by the diamonds.}\textbf{\footnotesize{} c}{\footnotesize{} Calculated
DOS involving several coupled SSPL in a double quantum dot model.
Here a SSPL is represented by a pair of dashed lines of the same color.
The positions of the levels and their coupling to the electrodes were
adjusted to provide best overall agreement with 2b. This simple model
captures many of the observed features and shows how ABS spectroscopy
allows the identification of the dot levels, and in particular of
their relative spin, without applying any magnetic field.}}

\end{figure}

\begin{figure}
\includegraphics[width=10cm]{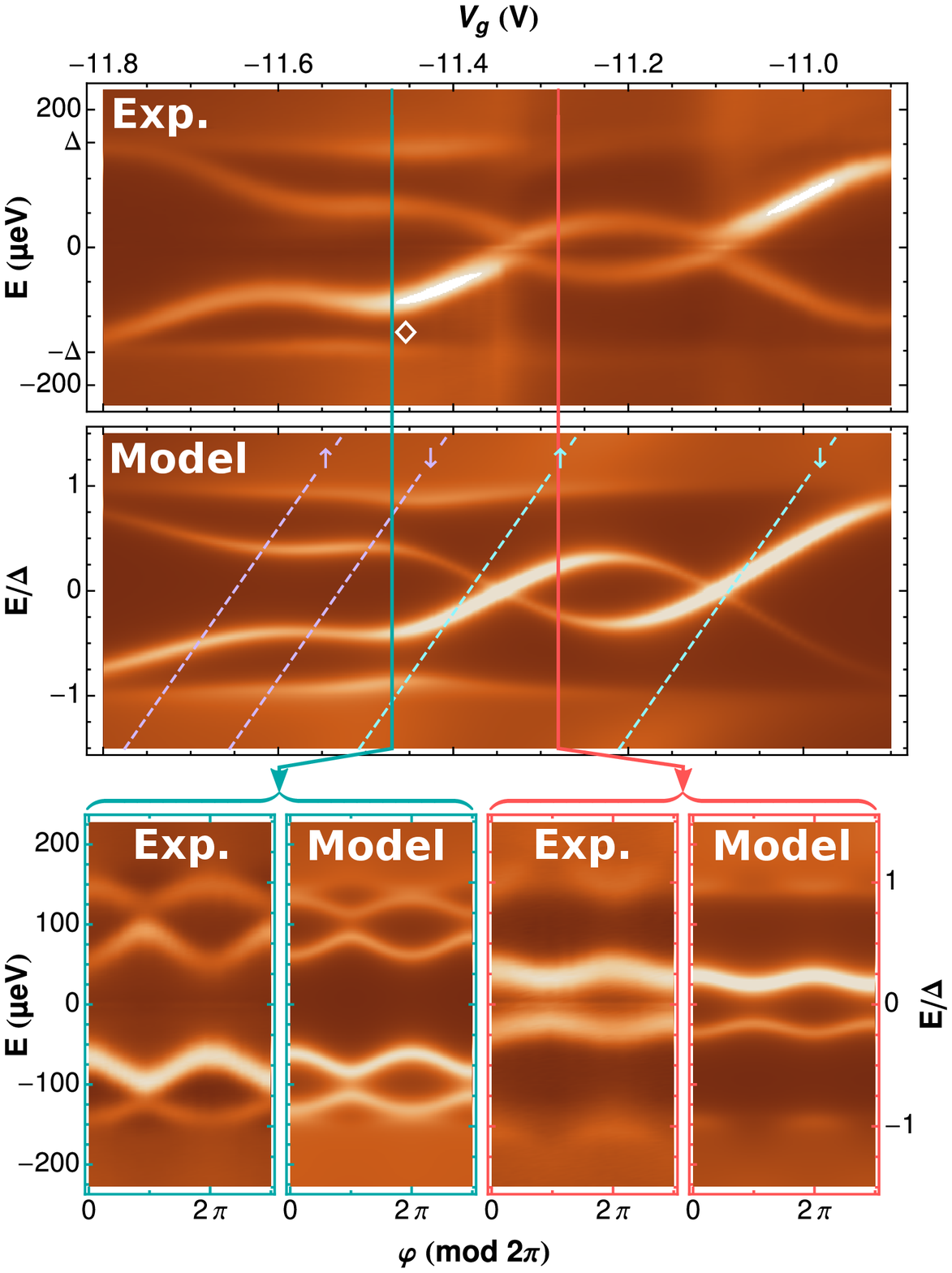}

\caption{{\footnotesize top panels: experimental deconvolved DOS as a function
of $V_{g}$ and predictions of our double quantum dot model, at phase
$\varphi=0$. Bottom pannels: experimental phase dependence taken
at gate voltages indicated by the plain coloured lines in the top
panels, and corresponding predictions of the model. All theoretical
panels use the same set of parameters; the corresponding level positions
are indicated by the dashed lines in the second pannel. The spin labelling
indicates only the relative orientations of the spins in these coupled
levels.}}

\end{figure}

Our sample is described in Figure 1. A CNT is well connected to two
superconducting metallic contacts 0.7 \textmu{}m apart, leaving enough
space to place a weakly-coupled tunnel electrode in between. The electrodes
are made of aluminum with a few nm of titanium as a sticking layer
(see SI for details); they become superconducting below $\sim1\,\mathrm{K}$.
The two outer contacts are reconnected, forming a loop. A magnetic
flux $\Phi$ threaded through the loop produces a superconducting
phase difference $\varphi=\frac{2e}{h}\Phi$ across the tube. By measuring
the differential conductance of the tunnel contact at low temperature
$(T\sim40\mathrm{\, mK})$ we observe (see Fig. 2a, 3a) well-defined
resonances inside the superconducting gap. The energies of these resonances
strongly depend on the voltage applied on the back-gate of the device,
and vary periodically with the phase difference accross the CNT, a
signature of ABS. From the raw measurement of the differential conductance
between the tunnel probe and the loop we can extract the density of
states (DOS) in the tube (see e.g. fig. 2b) through a straightforward
deconvolution procedure (see SI). Figure 2 shows the dependence of
the ABS spectrum on the flux in the loop at a fixed gate voltage.
By dc-biasing this device at a point which maximizes $\partial I/\partial\Phi$
(see Fig. 2a), it can be used as a SQUID magnetometer which combines
the advantages of Refs \cite{giazotto2010} and \cite{cleuziou2006}.
Being nanotube-based, our SQUID should be able to detect the reversal
of magnetic moments of only a few Bohr magnetons\cite{cleuziou2006}.
At the same time, the present device can be read out with a dc current
measurement (similar to \cite{giazotto2010}) and requires a single
gate voltage, making it easier to operate than Ref. \cite{cleuziou2006}.
The gate voltage dependence of the DOS shows a pattern of resonance
lines (Fig. 3b) which is more or less intricate depending on the strength
of the coupling to the leads (see SI).

We now show that the ABS observed in this device arise from the discrete
molecular levels in the CNT. For this we describe phenomenologically
our nanotube as a Quantum Dot (QD) coupled to superconducting leads
(See the SI for a detailed discussion of the model). The essential
physics of ABS in this system is already captured when one considers
a single orbital of the QD filled with either one or two electrons.
Due to the Pauli exclusion principle, these two electrons have opposite
spins and can thus be coupled by Andreev Reflection. Also, the doubly
occupied state is higher in energy by an effective charging energy
$\tilde{U}$ which can be determined from the experimental data. Hence,
the minimal effective model consists of a Spin-Split Pair of Levels
(SSPL) whose parameters are the splitting $\tilde{U}$, the mean position
$\overline{E}$ of the SSPL relative to the Fermi level (which is
controlled by the gate voltage $V_{g}$), and the coupling to the
leads (see Fig. S1a in SI). Previous theoretical work \cite{yoshioka2000,vecino2003}
has shown that there can be up to four ABS, symmetric (in position,
but not in intensity) about the Fermi Level. For sufficiently large
$\tilde{U}$ (respectively, $\overline{E}$), however, the two outer
(respectively, all) ABS merge with the continuum and are no longer
visible in the spectrum \cite{yoshioka2000,vecino2003,meng2009}.

We now discuss the dependence of the ABS energies on the gate voltage
$V_{g}$. The ABS appear as facing pairs of bell-shaped resonances
centred at $\overline{E}(V_{g})=0$ and with their bases resting against
opposite edges of the superconducting gap (see green dashed curves
in Fig. 3b.). For large enough $\tilde{U}$ the inner resonances cross
at the Fermi energy, forming a loop (Fig. 3b.). Such loops are a distinct
signature of SSPL in this model (spin-degenerate levels ($\tilde{U}=0$)
cannot give loops). Most of the features observed in Fig. 3b can be
identified as such pairs of bell-shaped resonances corresponding thus
to different SSPL in the nanotube.

Closer inspection reveals however that adjacent resonances are sometimes
coupled, forming avoided crossings (as indicated by $\diamondsuit$
symbols in Fig. 3b, 4), so that we need to consider the case where
two SSPL contribute simultaneously to the spectral properties within
the superconducting gap. For this, we extend the model to two serially-connected
QD each containing a SSPL, with a significant hopping term in between.
This model is fairly natural, given that the centre tunnel probe electrode
is likely to act as an efficient scatterer. The full description of
the model, the derivation of the retarded Green function from which
we obtain the spectral properties, and the parameters used to produce
the theoretical panels in Figures 3\&4 are detailed in the SI. Assuming
for simplicity that all states in the two dots are identically capacitively
coupled to the gate and that the couplings to the leads are independent
of $V_{g}$, we can locally reproduce most features of the gate-voltage
dependence of the DOS, and simultaneously the flux dependence at fixed
$V_{g}$ (see fig. 4). By summing contributions of independent SSPLs
and pairs of coupled SSPLs, (i.e. isolated orbitals and coupled pairs
of QD orbitals) we can also reproduce the observed dependence on an
extended $V_{g}$ range (See Fig. 3 b \& c, and discussion in the
SI).

Note that a single superconducting terminal is sufficient to induce
ABS in a QD (in which case, of course, there can be no supercurrent)
(see Refs. \cite{deacon2010,dirks2010}). Given this, and in light
of our analysis, we think that some features observed in Refs \cite{eichler2007,grove-rasmussen2009}
which were tentatively explained as out-of-equilibrium second order
AR can now be reinterpreted as equilibrium ABS spectroscopy on a QD
well connected to one superconducting lead, as in Refs. \cite{deacon2010,dirks2010},
with the second lead acting as a superconducting tunnel probe.

The agreement between experiment and theory in Figs. 3 and 4 shows
that ABS spectra constitute an entirely new spectroscopic tool for
QDs and CNTs. This spectroscopy provides extremely detailed information,
in particular about the relative spin state of the nanotube levels
without requiring high magnetic fields. Note that, in contrast to
the usual Coulomb blockade spectroscopy of QDs, the energy resolution
is here essentially independent of the temperature (as long as $k_{B}T\ll\Delta$)
and of the strength of the coupling to the leads. It should therefore
allow the exploration of the transition between the Fabry-Pérot (where
the Luttinger-Liquid physics is expected to play a role \cite{fazio1995,caux2002})
and the Coulomb blockade regimes in CNT. We also expect this new technique
to be able to provide key insights in cases where simple charge transport
measurements are not sufficient to fully probe the physics at work.
In particular, it should allow detailed investigation of the competition
between superconductivity and the Kondo effect \cite{glazman1989}
which arise for stronger couplings to the leads. Also, used in combination
with an in-plane magnetic field, it could also probe spin-orbit interactions
\cite{dellanna2007,dolcini2008,zazunov2009}. Finally it should be
emphasized that even while our phenomenological model successfully
describes the observed experimental data further theoretical work
is needed in order to establish a truly microscopic theory which should
predict the level splittings from the bare many-body Hamiltonian.

The information extracted from such spectroscopy may also help to
optimize Field Effect Transistors, SQUIDs or even Nano Electromechanical
devices based on nanotubes, by better understanding how current is
carried through the device. It could also be used for evaluating recently
proposed devices for quantum information processing such as entangled
electron pair generation by crossed Andreev reflection \cite{recher2001}
or ABS-based quantum bits \cite{zazunov2003}. Regarding the latter,
our observation of tunable ABS is heartening even though the measured
spectroscopic linewidth ($30-40\,\text{\textmu eV\,\ FWHM}$) seems
to question the feasability of such qubits (if it were intrinsic to
the sample, it would correspond to sub-ns coherence time). The present
linewidth is however likely to be caused simply by spurious noise
in the experimental setup. More investigations are needed in order
to assess the potential of nanotube ABS as qubits. 

To summarize, we have performed the first tunneling spectroscopy of
individually resolved ABS which provide a universal description for
the Josephson effect in weak links. The analysis of the ABS spectrum
constitutes a powerful and promising spectroscopic technique capable
of elucidating the electronic structure of CNT-based devices, including
those with well-coupled leads.

\begin{acknowledgments}
This work was partially supported by ANR project SEMAFAET, C'Nano
projet SPLONA, Spanish MICINN under contracts NAN2007-29366 and FIS2008-04209.

The authors gratefully acknowledge discussions with the Quantronics
group, Ophir Auslander, Juan Carlos Cuevas, Reinhold Egger, Milena
Grifoni, Takis Kontos, Hélène le Sueur, Alvaro Martín-Rodero, Pascal
Simon and Christoph Strunk.
\end{acknowledgments}

\section*{References}

\bibliographystyle{naturemag}
\bibliography{references}

\begin{thebibliography}{10}
\expandafter\ifx\csname url\endcsname\relax
  \def\url#1{\texttt{#1}}\fi
\expandafter\ifx\csname urlprefix\endcsname\relax\def\urlprefix{URL }\fi
\providecommand{\bibinfo}[2]{#2}
\providecommand{\eprint}[2][]{\url{#2}}

\bibitem{kasumov1999}
\bibinfo{author}{Kasumov, A.~Y.} \emph{et~al.}
\newblock \bibinfo{title}{Supercurrents through {Single-Walled} carbon
  nanotubes}.
\newblock \emph{\bibinfo{journal}{Science}} \textbf{\bibinfo{volume}{284}},
  \bibinfo{pages}{1508--1511} (\bibinfo{year}{1999}).

\bibitem{jarillo-herrero2006}
\bibinfo{author}{{Jarillo-Herrero}, P.}, \bibinfo{author}{{van Dam}, J.~A.} \&
  \bibinfo{author}{Kouwenhoven, L.~P.}
\newblock \bibinfo{title}{Quantum supercurrent transistors in carbon
  nanotubes}.
\newblock \emph{\bibinfo{journal}{Nature}} \textbf{\bibinfo{volume}{439}},
  \bibinfo{pages}{953--956} (\bibinfo{year}{2006}).

\bibitem{cleuziou2006}
\bibinfo{author}{Cleuziou, J.~P.}, \bibinfo{author}{Wernsdorfer, W.},
  \bibinfo{author}{Bouchiat, V.}, \bibinfo{author}{Ondarcuhu, T.} \&
  \bibinfo{author}{Monthioux, M.}
\newblock \bibinfo{title}{Carbon nanotube superconducting quantum interference
  device}.
\newblock \emph{\bibinfo{journal}{Nature Nanotechnology}}
  \textbf{\bibinfo{volume}{1}}, \bibinfo{pages}{53--59} (\bibinfo{year}{2006}).

\bibitem{pallecchi2008}
\bibinfo{author}{Pallecchi, E.}, \bibinfo{author}{Gaass, M.},
  \bibinfo{author}{Ryndyk, D.~A.} \& \bibinfo{author}{Strunk, C.}
\newblock \bibinfo{title}{Carbon nanotube {Josephson} junctions with {Nb}
  contacts}.
\newblock \emph{\bibinfo{journal}{Applied Physics Letters}}
  \textbf{\bibinfo{volume}{93}}, \bibinfo{pages}{072501--3}
  (\bibinfo{year}{2008}).

\bibitem{beenakker2004}
\bibinfo{author}{Beenakker, C. W.~J.}
\newblock \bibinfo{title}{Three "universal" mesoscopic {Josephson} effects}.
\newblock \emph{\bibinfo{journal}{cond-mat/0406127}}  (\bibinfo{year}{2004}).
\newblock \bibinfo{note}{Transport Phenomena in Mesoscopic Systems, edited by
  H. Fukuyama and T. Ando {(Springer,} Berlin, 1992)}.

\bibitem{recher2001}
\bibinfo{author}{Recher, P.}, \bibinfo{author}{Sukhorukov, E.~V.} \&
  \bibinfo{author}{Loss, D.}
\newblock \bibinfo{title}{{Andreev} tunneling, {Coulomb} blockade, and resonant
  transport of nonlocal spin-entangled electrons}.
\newblock \emph{\bibinfo{journal}{Physical Review B}}
  \textbf{\bibinfo{volume}{63}}, \bibinfo{pages}{165314}
  (\bibinfo{year}{2001}).

\bibitem{herrmann2010}
\bibinfo{author}{Herrmann, L.~G.} \emph{et~al.}
\newblock \bibinfo{title}{Carbon nanotubes as {Cooper-Pair} beam splitters}.
\newblock \emph{\bibinfo{journal}{Physical Review Letters}}
  \textbf{\bibinfo{volume}{104}}, \bibinfo{pages}{026801}
  (\bibinfo{year}{2010}).

\bibitem{zazunov2003}
\bibinfo{author}{Zazunov, A.}, \bibinfo{author}{Shumeiko, V.},
  \bibinfo{author}{Bratus{\textquoteright}, E.}, \bibinfo{author}{Lantz, J.} \&
  \bibinfo{author}{Wendin, G.}
\newblock \bibinfo{title}{{Andreev} level qubit}.
\newblock \emph{\bibinfo{journal}{Physical Review Letters}}
  \textbf{\bibinfo{volume}{90}} (\bibinfo{year}{2003}).

\bibitem{kulik1970}
\bibinfo{author}{Kulik, I.}
\newblock \bibinfo{title}{Macroscopic quantization and proximity effect in
  {S-N-S} junctions}.
\newblock \emph{\bibinfo{journal}{Soviet Physics {JETP-USSR}}}
  \textbf{\bibinfo{volume}{30}}, \bibinfo{pages}{944--\&}
  (\bibinfo{year}{1970}).

\bibitem{rocca2007}
\bibinfo{author}{{Della Rocca}, M.~L.} \emph{et~al.}
\newblock \bibinfo{title}{Measurement of the {Current-Phase} relation of
  superconducting atomic contacts}.
\newblock \emph{\bibinfo{journal}{Physical Review Letters}}
  \textbf{\bibinfo{volume}{99}}, \bibinfo{pages}{127005}
  (\bibinfo{year}{2007}).

\bibitem{fischer2007}
\bibinfo{author}{Fischer, {\O}.}, \bibinfo{author}{Kugler, M.},
  \bibinfo{author}{{Maggio-Aprile}, I.}, \bibinfo{author}{Berthod, C.} \&
  \bibinfo{author}{Renner, C.}
\newblock \bibinfo{title}{Scanning tunneling spectroscopy of high-temperature
  superconductors}.
\newblock \emph{\bibinfo{journal}{Reviews of Modern Physics}}
  \textbf{\bibinfo{volume}{79}}, \bibinfo{pages}{353} (\bibinfo{year}{2007}).

\bibitem{chen2009}
\bibinfo{author}{Chen, Y.}, \bibinfo{author}{Dirks, T.},
  \bibinfo{author}{{Al-Zoubi}, G.}, \bibinfo{author}{Birge, N.~O.} \&
  \bibinfo{author}{Mason, N.}
\newblock \bibinfo{title}{Nonequilibrium tunneling spectroscopy in carbon
  nanotubes}.
\newblock \emph{\bibinfo{journal}{Physical Review Letters}}
  \textbf{\bibinfo{volume}{102}}, \bibinfo{pages}{036804}
  (\bibinfo{year}{2009}).

\bibitem{bockrath1999}
\bibinfo{author}{Bockrath, M.} \emph{et~al.}
\newblock \bibinfo{title}{{Luttinger}-liquid behaviour in carbon nanotubes}.
\newblock \emph{\bibinfo{journal}{Nature}} \textbf{\bibinfo{volume}{397}},
  \bibinfo{pages}{598--601} (\bibinfo{year}{1999}).

\bibitem{buitelaar2002}
\bibinfo{author}{Buitelaar, M.~R.}, \bibinfo{author}{Nussbaumer, T.} \&
  \bibinfo{author}{Sch\"{o}nenberger, C.}
\newblock \bibinfo{title}{Quantum dot in the {Kondo} regime coupled to
  superconductors}.
\newblock \emph{\bibinfo{journal}{Physical Review Letters}}
  \textbf{\bibinfo{volume}{89}}, \bibinfo{pages}{256801}
  (\bibinfo{year}{2002}).

\bibitem{kuemmeth2008}
\bibinfo{author}{Kuemmeth, F.}, \bibinfo{author}{Ilani, S.},
  \bibinfo{author}{Ralph, D.~C.} \& \bibinfo{author}{{McEuen}, P.~L.}
\newblock \bibinfo{title}{Coupling of spin and orbital motion of electrons in
  carbon nanotubes}.
\newblock \emph{\bibinfo{journal}{Nature}} \textbf{\bibinfo{volume}{452}},
  \bibinfo{pages}{448--452} (\bibinfo{year}{2008}).

\bibitem{glazman1989}
\bibinfo{author}{Glazman, L.~I.} \& \bibinfo{author}{Matveev, K.~A.}
\newblock \bibinfo{title}{Resonant {Josephson} current through {{Kondo}}
  impurities in a tunnel barrier}.
\newblock \emph{\bibinfo{journal}{{JETP} Letters}}
  \textbf{\bibinfo{volume}{49}}, \bibinfo{pages}{659--662}
  (\bibinfo{year}{1989}).

\bibitem{spivak1991}
\bibinfo{author}{Spivak, B.~I.} \& \bibinfo{author}{Kivelson, S.~A.}
\newblock \bibinfo{title}{Negative local superfluid densities: The difference
  between dirty superconductors and dirty {B}ose liquids}.
\newblock \emph{\bibinfo{journal}{Physical Review B}}
  \textbf{\bibinfo{volume}{43}}, \bibinfo{pages}{3740} (\bibinfo{year}{1991}).

\bibitem{fazio1995}
\bibinfo{author}{Fazio, R.}, \bibinfo{author}{Hekking, F. W.~J.} \&
  \bibinfo{author}{Odintsov, A.~A.}
\newblock \bibinfo{title}{{Josephson} current through a {Luttinger} liquid}.
\newblock \emph{\bibinfo{journal}{Physical Review Letters}}
  \textbf{\bibinfo{volume}{74}}, \bibinfo{pages}{1843} (\bibinfo{year}{1995}).

\bibitem{caux2002}
\bibinfo{author}{Caux, J.}, \bibinfo{author}{Saleur, H.} \&
  \bibinfo{author}{Siano, F.}
\newblock \bibinfo{title}{{Josephson} current in {Luttinger}
  {Liquid-Superconductor} junctions}.
\newblock \emph{\bibinfo{journal}{Physical Review Letters}}
  \textbf{\bibinfo{volume}{88}}, \bibinfo{pages}{106402}
  (\bibinfo{year}{2002}).

\bibitem{dellanna2007}
\bibinfo{author}{{Dell'Anna}, L.}, \bibinfo{author}{Zazunov, A.},
  \bibinfo{author}{Egger, R.} \& \bibinfo{author}{Martin, T.}
\newblock \bibinfo{title}{{Josephson} current through a quantum dot with
  spin-orbit coupling}.
\newblock \emph{\bibinfo{journal}{Physical Review B}}
  \textbf{\bibinfo{volume}{75}}, \bibinfo{pages}{085305}
  (\bibinfo{year}{2007}).

\bibitem{dolcini2008}
\bibinfo{author}{Dolcini, F.} \& \bibinfo{author}{{Dell'Anna}, L.}
\newblock \bibinfo{title}{Multiple {Andreev} reflections in a quantum dot
  coupled to superconducting leads: Effect of spin-orbit coupling}.
\newblock \emph{\bibinfo{journal}{Physical Review B}}
  \textbf{\bibinfo{volume}{78}}, \bibinfo{pages}{024518}
  (\bibinfo{year}{2008}).

\bibitem{zazunov2009}
\bibinfo{author}{Zazunov, A.}, \bibinfo{author}{Egger, R.},
  \bibinfo{author}{Jonckheere, T.} \& \bibinfo{author}{Martin, T.}
\newblock \bibinfo{title}{Anomalous {Josephson} current through a {Spin-Orbit}
  coupled quantum dot}.
\newblock \emph{\bibinfo{journal}{Physical Review Letters}}
  \textbf{\bibinfo{volume}{103}}, \bibinfo{pages}{147004}
  (\bibinfo{year}{2009}).

\bibitem{giazotto2010}
\bibinfo{author}{Giazotto, F.}, \bibinfo{author}{Peltonen, J.~T.},
  \bibinfo{author}{Meschke, M.} \& \bibinfo{author}{Pekola, J.~P.}
\newblock \bibinfo{title}{Superconducting quantum interference proximity
  transistor}.
\newblock \emph{\bibinfo{journal}{Nature Physics}}
  \textbf{\bibinfo{volume}{6}}, \bibinfo{pages}{254--259}
  (\bibinfo{year}{2010}).

\bibitem{yoshioka2000}
\bibinfo{author}{Yoshioka, T.} \& \bibinfo{author}{Ohashi, Y.}
\newblock \bibinfo{title}{Numerical renormalization group studies on single
  impurity {Anderson} model in superconductivity: A unified treatment of
  magnetic, nonmagnetic impurities, and resonance scattering}.
\newblock \emph{\bibinfo{journal}{Journal of the Physical Society of Japan}}
  \textbf{\bibinfo{volume}{69}}, \bibinfo{pages}{1812--1823}
  (\bibinfo{year}{2000}).

\bibitem{vecino2003}
\bibinfo{author}{Vecino, E.}, \bibinfo{author}{{Mart\'{i}n-Rodero}, A.} \&
  \bibinfo{author}{{Levy Yeyati}, A.}
\newblock \bibinfo{title}{{Josephson} current through a correlated quantum
  level: {Andreev} states and pi junction behavior}.
\newblock \emph{\bibinfo{journal}{Physical Review B}}
  \textbf{\bibinfo{volume}{68}}, \bibinfo{pages}{035105}
  (\bibinfo{year}{2003}).

\bibitem{meng2009}
\bibinfo{author}{Meng, T.}, \bibinfo{author}{Florens, S.} \&
  \bibinfo{author}{Simon, P.}
\newblock \bibinfo{title}{Self-consistent description of {Andreev} bound states
  in {Josephson} quantum dot devices}.
\newblock \emph{\bibinfo{journal}{Physical Review B {(Condensed} Matter and
  Materials Physics)}} \textbf{\bibinfo{volume}{79}},
  \bibinfo{pages}{224521--10} (\bibinfo{year}{2009}).

\bibitem{deacon2010}
\bibinfo{author}{Deacon, R.~S.} \emph{et~al.}
\newblock \bibinfo{title}{Tunneling spectroscopy of {Andreev} energy levels in
  a quantum dot coupled to a superconductor}.
\newblock \emph{\bibinfo{journal}{Physical Review Letters}}
  \textbf{\bibinfo{volume}{104}}, \bibinfo{pages}{076805}
  (\bibinfo{year}{2010}).

\bibitem{dirks2010}
\bibinfo{author}{Dirks, T.} \emph{et~al.}
\newblock \bibinfo{title}{{A}ndreev bound state spectroscopy in a graphene
  quantum dot}.
\newblock \emph{\bibinfo{journal}{arXiv:1005.2749}}  (\bibinfo{year}{2010}).
\newblock \urlprefix\url{http://arxiv.org/abs/1005.2749}.

\bibitem{eichler2007}
\bibinfo{author}{Eichler, A.} \emph{et~al.}
\newblock \bibinfo{title}{{Even-Odd} effect in {Andreev} transport through a
  carbon nanotube quantum dot}.
\newblock \emph{\bibinfo{journal}{Physical Review Letters}}
  \textbf{\bibinfo{volume}{99}}, \bibinfo{pages}{126602--4}
  (\bibinfo{year}{2007}).

\bibitem{grove-rasmussen2009}
\bibinfo{author}{{Grove-Rasmussen}, K.} \emph{et~al.}
\newblock \bibinfo{title}{Superconductivity-enhanced bias spectroscopy in
  carbon nanotube quantum dots}.
\newblock \emph{\bibinfo{journal}{Physical Review B {(Condensed} Matter and
  Materials Physics)}} \textbf{\bibinfo{volume}{79}},
  \bibinfo{pages}{134518--5} (\bibinfo{year}{2009}).

\end{thebibliography}

\end{document}